\documentstyle[amssymb,preprint,aps]{revtex}

\begin{document}
\title{Contextual Realization of the Universal Quantum Cloning Machine and of the
Universal-NOT gate by Quantum Injected Optical Parametric Amplification}
\author{D. Pelliccia, V. Schettini, F. Sciarrino, C. Sias and F. De Martini}
\address{Dipartimento di Fisica and \\
Istituto Nazionale per la Fisica della Materia\\
Universit\`{a} di Roma ''La Sapienza'', Roma, 00185 - Italy}
\maketitle

\begin{abstract}
A simultaneous, contextual experimental demonstration of the two processes
of {\it cloning} an input qubit $|\Psi >$and of \ {\it flipping} it into the
orthogonal qubit $|\Psi ^{\perp }>\ $is\ reported. The adopted experimental
apparatus, a Quantum-Injected Optical Parametric Amplifier (QIOPA) is
transformed simultaneously into a Universal Optimal Quantum Cloning Machine
(UOQCM) and into a Universal NOT\ quantum-information gate. The two
processes, indeed {\it forbidden} in their {\it exact} form for fundamental
quantum limitations, will be found to be {\it universal} and {\it optimal},
i.e. the measured {\it fidelity} of both processes $F<1$ will be found close
to the limit values evaluated by quantum theory.\ A\ contextual theoretical
and experimental investigation of these processes, which may represent the
basic difference between the classical and the quantum worlds, can reveal in
a unifying manner the detailed structure of quantum information. It may also
enlighten the yet little explored interconnections of \ fundamental
axiomatic properties within the deep structure of quantum mechanics. PACS
numbers: 03.67.-a, 03.65.Ta, 03.65.Ud
\end{abstract}

\section{Introduction}

Classical information is represented by {\it bits} which can be either 0 or
1. Quantum information is represented by {\it quantum-bits,} or ''qubits''
which are two-dimensional quantum systems. A qubit unlike a classical bit
can exist in a state $|\Psi >$ that is a superposition of two orthogonal
basis states $\{|\uparrow >;|\downarrow >\}$, i.e. $|\Psi >=\widetilde{%
\alpha }|\uparrow >+\widetilde{\beta }|\downarrow >$. The fact that qubits
can exist in these superposition states gives quantum information unusual
properties. Specifically, information encoded in quantum system has to obey
rules of quantum physics which impose strict bounds on possible
manipulations with quantum information. The common denominator of these
bounds is that all quantum-mechanical transformations have to be represented
by completely positive (CP)\ maps \cite{1} which in turn impose a constraint
on the fidelity of quantum-mechanical measurements. That is, an unknown
state of a qubit cannot be precisely determined (or reconstructed) from a
measurement performed on a finite ensemble of identically prepared qubits 
\cite{2,3,4}. In particular, the mean fidelity of the best possible
(optimal) state estimation strategy based on the measurement of $N$
identically prepared qubits is $F=(N+1)/(N+2)$. One of the obvious
consequences of this bound on the fidelity of estimation is that unknown
states of quantum systems cannot be cloned, viz. copied perfectly \cite{5},
namely the perfect cloning map of the form $|\Psi >\Longrightarrow |\Psi
>|\Psi >$ is not permitted by the rules of quantum mechanics. Certainly if
this would be possible, then one would be able to violate the bound on the
fidelity of estimation. Moreover, this possibility would trigger more
dramatic changes in the present picture of the physical world, e.g. it would
be possible to utilize quantum nonlocality for superluminal signaling \cite
{6,7,8}. Another map which cannot be performed perfectly on an {\it unknown}
state of a qubit is the {\it spin-flip} or the universal-NOT , i.e. the
operation $|\Psi >\Longrightarrow |\Psi ^{\perp }>$, where the state $|\Psi
^{\perp }>$ is orthogonal to the original $|\Psi >$ \cite{3,9}.
Spin-flipping is indeed an {\it anti-unitary}, i.e. time reversal map $%
T=i\sigma _{y}K$\ which realizes for any input qubit the {\it inversion}
over the Bloch Sphere, as shown in Figure1. Precisely, the {\it %
phase-conjugation} operator $K\;$is responsible for anti-unitarity since $%
K|\Psi >=|\Psi ^{\ast }>$, the complex conjugate of $|\Psi >$.

In spite of the fact that some quantum-mechanical transformations on unknown
states of qubits cannot be performed perfectly one still may ask that within
the given structure of quantum theory, i.e. which is {\it linear} and where
all maps are CP\ maps \cite{1,10}, what are the best possible approximations
of these maps. Namely, in the present context, what is the {\it optimal}
Universal Cloning and the {\it optimal} Universal-NOT (U-NOT) gate. The {\it %
universality} condition is required to ensure that all {\it unknown} inputs
states, i.e. all points on the Bloch Sphere, are transformed with the same ''%
{\it Quantum Efficiency}'' $(QE)$\ viz. ''{\it fidelity''}. Investigation of
these universal optimal transformations, which are also called {\it %
universal quantum machines} \cite{11} is extremely important since it
reveals bounds on optimal manipulations with quantum information.
Consequently, in recent years theoretical investigation on\ the Universal
Optimal Quantum Cloning Machine (UOQCM), and on the U-NOT gate\ has been
very thorough. In spite all the success in the theoretical analysis of the
bounds on optimal manipulations with individual qubits, it is extremely
difficult to realize experimentally universal quantum machines. In the
domain of quantum optics this is possible by associating a cloning machine
with a photon amplification process, e.g. involving inverted atoms in a
laser amplifier or a nonlinear (NL) medium in a {\it quantum-injected} {\it %
optical} {\it parametric amplifier} (QIOPA) \cite{12}. This can be done in
virtue of the existing isomorphism between any logic state of a qubit and
the polarization state of \ the photon {\it qubit}. In the case of the mode
non degenerate QIOPA \cite{12} it is generally supposed that $N$ photons,
prepared identically in an arbitrary quantum state (qubit) $|\Psi >$ of {\it %
linear-} {\it polarization} ($\overrightarrow{{\bf \pi }}$) are injected
into the amplifier on the input mode $k_{1}$.\ The amplifier then generates
on the same output ''cloning mode''\ (C) $M>N$ copies, or ''clones'' of the
input qubit $|\Psi >$. Correspondingly, the OPA\ amplifier generates on the
output ''anticloning mode''\ (AC), $M-N$ states $|\Psi ^{\perp }>$, thus
realizing a quantum NOT\ gate.

The work is organized as follows. In Section II the general theory of
quantum injected amplification is reviewed with emphasis on the dynamical
conditions apt to ensure, in the present context, the {\it universality}
property of the device, i.e. implying equal quantum efficiencies for {\it any%
} arbitrary input $\overrightarrow{\pi }$-state. Furthermore, the
multiparticle superposition state (or {\it Schroedinger Cat} state)\ of the
amplified field on modes C and AC will be investigated on the basis of the
Wigner function of the state. In Section III the achievement of the
universality property will be tested experimentally by injection of \ a
''classical'', Glauber {\it coherent} field. Details of the overall
experiment will be given in this Section. In Section IV\ the theory of the 
{\it optimal} no-cloning process and NOT gate will be outlined in a unitary
and consistent fashion\ for the $N=1,M=2$ case and applied to the QIOPA
scheme operating in a $\overrightarrow{\pi }-entangled$ configuration \cite
{12}. In Section V the experimental demonstration for both optimal processes
are reported, the values of the corresponding ''fidelity'' evaluated and
found in good agreement with the theoretical ones \cite{4,13}. For the sake
of completeness, in Section V the {\it universality} of the QIOPA apparatus
will tested again under the {\it quantum injection} of a single qubit: $N=1$%
. In conclusion, Section VI will be devoted to a theoretical discussion over
the inner connections existing between the two basic quantum information
processes, here investigated contextually by the same overall dynamical
process.

\section{Quantum Injected Optical Parametric Amplifier}

In the present work the quantum information carriers or {\it qubits} are
assumed to represent states of polarization $(\overrightarrow{\pi })$, and
QE is expressed by the OPA\ parametric ''gain'': $g$. Let us investigate
theoretically the dynamics of the QIOPA\ apparatus making reference to
Figure 2 and to Ref.12. The active element of the device is a Type II
nonlinear (NL) crystal slab operating in non-collinear configuration. In
these conditions the overall amplification taking place over the coupled
electromagnetic (e.m.)\ modes $k_{j}$ $(j=1,2)$ is contributed by two equal
and {\it independent} parametric amplifiers: OPA $A$ and OPA $\ A^{^{\prime
}}\;$inducing uncorrelated unitary transformations respectively on two
couples of (time) t-dependent field operators $\hat{a}_{1}(t)\equiv \hat{a}%
_{1h}(t),\hat{a}_{2}(t)\equiv \hat{a}_{2v}(t)$ and $\hat{a}_{1}^{\prime
}(t)\equiv \hat{a}_{1v}(t),\hat{a}_{2}^{\prime }(t)\equiv \hat{a}_{2h}(t)$
acting on the output modes $j=1,2$ along the horizontal (H{\bf ) }and{\bf \ }%
vertical (V) directions in the $\overrightarrow{\pi }$ plane. The
interaction Hamiltonian may be expressed in the general form:

\begin{equation}
\widehat{H}_{I}=i\hslash \chi \lbrack \hat{A}-e^{i\Phi }\hat{A}^{^{\prime
}}]+h.c.
\end{equation}
where: $\hat{A}\equiv $ $\hat{a}_{1}(t)\hat{a}_{2}(t)$, $\hat{A}^{^{\prime
}}\equiv \hat{a}_{1}^{^{\prime }}(t)\hat{a}_{2}^{^{\prime }}(t)$, and $%
g\equiv \chi t$ is a real number expressing the amplification gain
proportional to and the NL coupling term $\chi .$ The dynamics of OPA $A$
and OPA $A^{^{\prime }}$ is expressed correspondingly by the mutually
commuting unitary squeeze operators: $\widehat{U}_{A}(t)=exp[-g(\hat{A}%
^{\dagger }-\hat{A})]$ and $\widehat{U}_{A^{^{\prime }}}(t)=exp[g(e^{-i\Phi }%
\hat{A}^{^{\prime }\dagger }-e^{i\Phi }\hat{A}^{^{\prime }})]$ implying the
following Bogoliubov transformations \cite{13}: \newline
\begin{equation}
{\hat{a}_{1}(t) \atopwithdelims[] \hat{a}_{2}(t)^{\dagger }}%
=\left[ 
\begin{array}{cc}
C & S \\ 
S & C
\end{array}
\right] 
{\hat{a}_{1} \atopwithdelims[] \hat{a}_{2}^{\dagger }}%
;\ \ 
{\hat{a}_{1}^{\prime }(t) \atopwithdelims[] \hat{a}_{2}^{\prime }(t)^{\dagger }}%
=\left[ 
\begin{array}{cc}
C & \widetilde{S} \\ 
\widetilde{S}^{\ast } & C
\end{array}
\right] 
{\hat{a}_{1}^{\prime } \atopwithdelims[] \hat{a}_{2}^{\prime \dagger }}%
\end{equation}
\newline
where: $C\equiv cosh(g)$, $S\equiv sinh(g)$, $\widetilde{S}\equiv \epsilon S$%
, $\Gamma \equiv S/C$, $\widetilde{\Gamma }=\epsilon \Gamma $,$\ \varepsilon
\equiv -e^{-i\Phi }\ $and $\Phi $\ is an externally adjustable {\it %
intrinsic phase }existing{\it \ }between the $A$ and $A^{\prime }\ $OPA
devices. These transformations imply the time invariance of the interaction
Hamiltonian and of the field commutors, i.e. $\widehat{H}_{I}(t)=\widehat{H}%
_{I}(0)$, $[\hat{a}_{i}(t),\hat{a}_{j}^{\dagger }(t)]=[\hat{a}_{i},\hat{a}%
_{j}^{\dagger }]=\delta _{ij}$, $[\hat{a}_{i}^{^{\prime }}(t),\hat{a}%
_{j}^{^{\prime }\dagger }(t)]=[\hat{a}_{i}^{^{\prime }},\hat{a}%
_{j}^{^{\prime }\dagger }]=\delta _{ij}$, $[\hat{a}_{i}^{^{\prime }}(t),\hat{%
a}_{j}^{\dagger }(t)]=0$ being $i,j=1,2$ and $\hat{a}_{i,j}\equiv \hat{a}%
_{i,j}(0)$ , $\hat{a}_{i,j}^{^{\prime }}\equiv \hat{a}_{i,j}^{^{\prime }}(0)$
the input fields at the initial time $t=0$ i.e. before the OPA interaction.
The evolution in the Schroedinger picture of the state acted upon by the
OPA\ system is determined by the overall operator $\widehat{U}=\widehat{U}%
_{A}\otimes \widehat{U}_{A^{\prime }}$ expressed in terms of the operators
evaluated at $t=0$: 
\begin{equation}
\widehat{U}_{A}=\exp [g(\widehat{\sigma }_{+}+\widehat{\sigma }_{-})];\ 
\widehat{U}_{A^{\prime }}=\exp [g(\widehat{\sigma }_{+}^{\prime }+\widehat{%
\sigma }_{-}^{\prime })];
\end{equation}
\newline
by adopting the definitions: $\widehat{\sigma }_{+}$=$-\hat{A}^{\dagger }$,$%
\ \widehat{\sigma }_{-}$=$\hat{A},\ \widehat{\sigma }_{z}$=$(1+\hat{a}%
_{1}^{\dagger }\hat{a}_{1}+\hat{a}_{2}^{\dagger }\hat{a}_{2})\equiv (1+%
\widehat{n}_{1}+\widehat{n}_{2})$,$\ \widehat{\sigma }_{+}^{\prime
}=-\epsilon \hat{A}^{\prime \dagger }$,$\ \widehat{\sigma }_{-}^{\prime
}=\epsilon ^{\ast }\hat{A}^{\prime }$;$\;\widehat{\sigma }_{z}=(1+\widehat{n}%
_{1}^{\prime }+\widehat{n}_{2}^{\prime })$. In virtue of Eq.(2) the
following commutation properties for the sets of the $\widehat{\sigma }$ and 
$\widehat{\sigma }^{\prime }$pseudo-spin operators hold: $[\widehat{\sigma }%
_{+},\widehat{\sigma }_{-}]=\widehat{\sigma }_{z}\ $; $[\widehat{\sigma }%
_{z}\ ,\widehat{\sigma }_{\pm }]=\pm 2\widehat{\sigma }_{\pm }\ $and: $[%
\widehat{\sigma }_{+}^{\prime },\widehat{\sigma }_{-}^{\prime }]=\sigma
_{z}^{\prime }\ $;$[\widehat{\sigma }_{z}^{\prime },\widehat{\sigma }_{\pm
}^{\prime }]=\pm 2\widehat{\sigma }_{\pm }^{\prime }$, $[\widehat{\sigma }%
_{\pm },\widehat{\sigma }_{\mp }^{\prime }]=[\widehat{\sigma }_{\pm },%
\widehat{\sigma }_{z}^{\prime }]=0$. By further adopting the definitions: $%
\widehat{\sigma }_{x}\equiv 2^{-1/2}(\widehat{\sigma }_{+}+i\widehat{\sigma }%
_{-})$, $\widehat{\sigma }_{y}\equiv 2^{-1/2}(\widehat{\sigma }_{+}-i%
\widehat{\sigma }_{-})$, the following relevant commutors: $[\widehat{\sigma 
}_{x},\widehat{\sigma }_{y}]=-i\widehat{\sigma }_{z}$, $[\widehat{\sigma }%
_{x}^{\prime },\widehat{\sigma }_{y}^{\prime }]=-i\widehat{\sigma }%
_{z}^{\prime }$ are recognized as those belonging to the symmetry group
SU(1,1) \cite{14}. The output field may be expressed, in virtue of an
appropriate ''operator disentangling theorem'' in the following form \cite
{12,15}: 
\begin{equation}
\left| \Psi \right\rangle =\{\exp \Gamma (\widehat{\sigma }_{+}+\widehat{%
\sigma }_{+}^{\prime })\times \exp [-\ln C(\widehat{\sigma }_{z}+\widehat{%
\sigma }_{z}^{\prime })]\times \exp \Gamma (\widehat{\sigma }_{-}+\widehat{%
\sigma }_{-}^{\prime })\}\left| \Psi \right\rangle _{IN}.
\end{equation}
Take as input state into the QIOPA\ system the general {\it qubit} : $\left|
\Psi \right\rangle _{IN}\equiv (\widetilde{\alpha }\left| \Psi \right\rangle
_{IN}^{\alpha }+\widetilde{\beta }\left| \Psi \right\rangle _{IN}^{\beta })$%
, $\left| \widetilde{\alpha }\right| ^{2}+\left| \widetilde{\beta }\right|
^{2}=1$, defined in the $2\times 2$-dimensional Hilbert space of
polarizations $(\overrightarrow{\pi })$ on the $2$ interacting modes $k_{1}\ 
$and $k_{2}$ with basis vectors: $\left| \Psi \right\rangle _{IN}^{\alpha
}=\left| 1\right\rangle _{1h}\left| 0\right\rangle _{1v}\left|
0\right\rangle _{2h}\left| 0\right\rangle _{2v}\equiv \left|
1,0,0,0\right\rangle $, $\left| \Psi \right\rangle _{IN}^{\beta }=\left|
0,1,0,0\right\rangle $. Here the general product state $\left|
x\right\rangle _{1h}\otimes \left| y\right\rangle _{1v}\otimes \left|
z\right\rangle _{2h}\otimes \left| t\right\rangle _{2v}$ has been, and shall
be henceforth, expressed by the shorthand: $\left| x,y,z,t\right\rangle $.
In virtue of the general \ {\it information preserving \ }property of any
nonlinear (NL) transformation of parametric type, the output state is again
expressed by a ''{\it massive qubit}'': $\left| \Psi \right\rangle \equiv (%
\widetilde{\alpha }\left| \Psi \right\rangle ^{\alpha }+\widetilde{\beta }%
\left| \Psi \right\rangle ^{\beta })\ $\cite{16}. This (pure)\ state is
indeed a\ \ {\it Schroedinger-Cat (S-Cat) }state implying the quantum
superposition of the orthonormal multiparticle states \cite{12}: 
\begin{equation}
\left| \Psi \right\rangle ^{\alpha }\equiv C^{-3}\sum\limits_{i,\
j=0}^{\infty }(-\Gamma ^{i+j})\epsilon ^{j}\sqrt{i+1}\ \left|
i+1,j,j,i\right\rangle ;\ \ \left| \Psi \right\rangle ^{\beta }\equiv
C^{-3}\sum\limits_{i,\ j=0}^{\infty }(-\Gamma ^{i+j})\varepsilon ^{j}\sqrt{%
j+1\ }\left| i,j+1,j,i\right\rangle
\end{equation}
Consider the density operator $\rho \equiv \left| \Psi \right\rangle
\left\langle \Psi \right| \ $and his reductions over the $\overrightarrow{%
\pi }-vector$ spaces relative to the spatial modes $k_{1}$and $k_{2}$: $\
\rho _{1}=Tr_{2}\rho \ $;$\ \rho _{2}=Tr_{1}\rho $. These ones may be
expanded as a weighted superpositions of $p-square$ matrices of order $%
p=(n+2)$, the relative weight $\Gamma ^{2}=\tanh ^{2}g\ $of each two
successive matrices being determined by the parametric ''gain''. $\Gamma
^{2}=0$ for $g=0$ and approaches asymptotically the unit value for large $g$%
.\ In turn, the $p-square$ matrices may be expressed as sum of $2\times 2$
matrices as shown by the following expressions: 
\begin{equation}
\rho _{1}=C^{-6}\sum\limits_{n=0}^{\infty }\Gamma ^{2n}\times
\sum\limits_{i=\ 0}^{n}\left[ 
\begin{array}{cc}
\left| \widetilde{\beta }\right| ^{2}(n-i+1) & \widetilde{\alpha }^{\ast }%
\widetilde{\beta }\sqrt{(i+1)(n-i+1)} \\ 
\widetilde{\alpha }\widetilde{\beta }^{\ast }\sqrt{(i+1)(n-i+1)} & \left| 
\widetilde{\alpha }\right| ^{2}(i+1)
\end{array}
\right]
\end{equation}
\ written in terms of the Fock basis: $\left\{ \left| i\right\rangle
_{1h}\left| n-i+1\right\rangle _{1v}\text{;\ }\left| i+1\right\rangle
_{1h}\left| n-i\right\rangle _{1v}\right\} $. Correspondingly: 
\begin{equation}
\rho _{2}=C^{-6}\sum\limits_{n=0}^{\infty }\Gamma ^{2n}\times
\sum\limits_{i=0}^{n+1}\left[ 
\begin{array}{cc}
\left| \widetilde{\beta }\right| ^{2}(n-i+1) & \epsilon ^{\ast }\widetilde{%
\alpha }^{\ast }\widetilde{\beta }\sqrt{(n-i+1)i} \\ 
\epsilon \widetilde{\alpha }\widetilde{\beta }^{\ast }\sqrt{(n-i+1)i} & 
\left| \widetilde{\alpha }\right| ^{2}i
\end{array}
\right]
\end{equation}
in terms of the Fock basis:$\left\{ \left| n-i\right\rangle _{2h}\left|
i\right\rangle _{2v}\text{;\ }\left| n-i+1\right\rangle _{2h}\left|
i-1\right\rangle _{2v}\right\} $. \ Interestingly enough, the value $n$ of
the sum indices appearing in Eqs. 6 and 7 coincides with the {\it number} of
photon pairs generated by the QIOPA\ amplification. Note that all the $%
2\times 2$ matrices in Eqs. 6, 7 and then the $p-square$ matrices resulting
from their sums over the $i-index$, are non-diagonal as implied by the
quantum superposition property of any {\it Schroedinger-Cat} state. Note
also that the OPA\ {\it intrinsic phase }$\Phi \ $only affects the {\it %
anticloning channel }(AC), i.e. the mode $k_{2}$.

{\large Wigner Function.} In order to inspect at a deeper lever the {\it %
S-Cat} condition, consider the Wigner function of the output field for the
QIOPA apparatus shown in Figure 1. Evaluate first the symmetrically-ordered
characteristic function of the set of complex variables $(\eta _{j},\eta 
{\ast  \atop j}%
,\xi _{j},\xi 
{\ast  \atop j}%
)\equiv \left\{ \eta ,\xi \right\} $, $(j=1,2)$: $\chi _{_{S}}\left\{ \eta
,\xi \right\} \equiv \left\langle \Psi _{0}\right| D[\eta _{1}(t)]D[\eta
_{2}(t)]D[\xi _{1}(t)]D[\xi _{2}(t)]\left| \Psi _{0}\right\rangle $
expressed in terms of the {\it displacement }operators: $D[\eta
_{j}(t)]\equiv $ $\exp [\eta _{j}(t)\hat{a}_{j}^{\dagger }-\eta _{j}^{\ast
}(t)\hat{a}_{j}]$, ${\it \ }D[\xi _{j}(t)]$ $\ \equiv \exp [\xi _{j}(t)%
\widehat{a}_{j}^{\prime \dagger }-\xi _{j}^{\ast }(t)\widehat{a}_{j}^{\prime
}]$ \ where: $\eta _{1}(t)\equiv (\eta _{1}C-\eta 
{\ast  \atop 2}%
S)$; $\eta _{2}(t)\equiv (\eta _{2}C-\eta 
{\ast  \atop 1}%
S)$; $\xi _{1}(t)\equiv (\xi _{1}C-\xi 
{\ast  \atop 2}%
S)$; $\xi _{2}(t)\equiv (\xi _{2}C-\xi 
{\ast  \atop 1}%
S)$. The Wigner function, expressed in terms of the corresponding complex
phase-space variables $(\alpha _{j}$, $\alpha _{j}^{\ast }$, $\beta _{j}$, $%
\beta _{j}^{\ast })\equiv \left\{ \alpha ,\beta \right\} $ is the
eight-dimensional Fourier transform of $\chi _{_{S}}\left\{ \eta ,\xi
\right\} $, namely: 
\begin{equation}
W\left\{ \alpha ,\beta \right\} =\pi ^{-8}\int \int \int \int d^{2}\eta
_{1}d^{2}\eta _{2}d^{2}\xi _{1}d^{2}\xi _{2}\chi _{_{S}}\left\{ \eta ,\xi
\right\} \exp \sum_{j}[(\eta 
{\ast  \atop j}%
\alpha _{j}+\xi 
{\ast  \atop j}%
\beta _{j})-c.c.]
\end{equation}
where $d^{2}\eta _{j}\equiv d\eta _{j}d\eta 
{\ast  \atop j}%
$, etc. By a lengthy application of operator algebra and integral calculus,
reported in \cite{17}, we could evaluate analytically in closed form either $%
\chi _{_{S}}\left\{ \eta ,\xi \right\} $\ and $W\left\{ \alpha ,\beta
\right\} $.The {\it exact} expression of the Wigner function is found: 
\begin{equation}
W\left\{ \alpha ,\beta \right\} =-\overline{W_{A}}\left\{ \alpha \right\} 
{\it \ }\overline{W}_{A^{\prime }}\left\{ \beta \right\} \left[ 1-\left|
e^{i\Phi }\Delta _{A}\left\{ \alpha \right\} +\Delta _{A^{\prime }}\left\{
\beta \right\} \right| ^{2}\right]
\end{equation}
where $\Delta _{A}\left\{ \alpha \right\} \equiv 2^{-%
{\frac12}%
}(\gamma _{A+}-i\gamma _{A-})$, $\Delta _{A^{\prime }}\left\{ \beta \right\}
\equiv 2^{-%
{\frac12}%
}(\gamma _{A^{\prime }+}-i\gamma _{A^{\prime }-})$ are expressed in terms of
the squeezed variables{\it : }$\gamma _{A+}\equiv (\alpha _{1}+\alpha
_{2}^{\ast })e^{-g}$; $\gamma _{A-}\equiv i(\alpha _{1}-\alpha _{2}^{\ast
})e^{+g}$; $\gamma _{A^{\prime }+}\equiv (\beta _{1}+\beta _{2}^{\ast
})e^{-g}$; $\gamma _{A^{\prime }-}\equiv i(\beta _{1}-\beta _{2}^{\ast
})e^{+g}$. The Wigner functions $\overline{W_{A}}\left\{ \alpha \right\}
\equiv 4\pi ^{-2}\exp \left( -\left[ \left| \gamma _{A+}\right| ^{2}+\left|
\gamma _{A-}\right| ^{2}\right] \right) $; $\overline{W_{A^{\prime }}}%
\left\{ \beta \right\} \equiv 4\pi ^{-2}\exp \left( -\left[ \left| \gamma
_{A^{\prime }+}\right| ^{2}+\left| \gamma _{A^{\prime }-}\right| ^{2}\right]
\right) $ definite positive over the 4 - dimensional spaces $\left\{ \alpha
\right\} $ and $\left\{ \beta \right\} $represent the effect of
squeezed-vacuum, i.e. emitted respectively by OPA$_{A}$\ and OPA$_{A^{\prime
}}$ in absence of any injection. Inspection of Equation 9 shows that
precisely the quantum superposition character of the injected state $\left|
\Psi \right\rangle _{IN}$ determines the dynamical quantum superposition of
the devices $OPA_{A}\ $and $OPA_{A^{\prime }},$~the ones that otherwise act
as {\it uncoupled} and{\it \ independent\ }objects. In other words, the
quasi-probability functions $\overline{W_{A}}\left\{ \alpha \right\} $,{\it %
\ }$\overline{W_{A^{\prime }}}\left\{ \beta \right\} $corresponding to the
two macrostates $\left| \Psi \right\rangle ^{\alpha }\ $and $\left| \Psi
\right\rangle ^{\beta }\ $in absence of quantum superposition are defined in
two totally separated and independent spaces $\left\{ \alpha \right\} $ and$%
\ \left\{ \beta \right\} $. The link between these spaces is provided by the
superposition term in Equation 9: $2Re[e^{i\Phi }\Delta _{A}\left\{ \alpha
\right\} \Delta _{A^{\prime }}^{\ast }\left\{ \beta \right\} ]$. This term
provides precisely the first-order quantum interference of the macrostates $%
\left| \Psi \right\rangle ^{\alpha }\ $and $\left| \Psi \right\rangle
^{\beta }$. In addition, and most important, Equation 9 shows the non
definite positivity{\it \ }of $W\left\{ \alpha ,\beta \right\} $ over its
definition space. This assures the overall quantum character of our
multiparticle, injected amplification scheme \cite{12,18}.

{\large Universality}. A necessary common property of the QIOPA\ system in
the context of the present work is its {\it universality} (U), i.e. implying
the {\it same} quantum efficiency (QE) of the amplifying apparatus for any
input, unknown qubit. That is, for a qubit spanning the entire Bloch sphere.
In our experiment the qubits are assumed to represent states of polarization 
$(\overrightarrow{\pi })$, as said, and QE is expressed by the QIOPA\
parametric {\it gain}: $g$. We shall find that universality implies an
important symmetry property, namely the invariance of the coupling
Hamiltonian $H_{int}$ under simultaneous general $SU(2)$ transformations on
the polarization $\overrightarrow{\pi }$ on the spatial\ modes $k_{j}$ $%
(j=1,2)$ \cite{13,19}. Assume that under a simultaneous general $%
\overrightarrow{\pi }$ rotation $R\;$of $\overrightarrow{\pi }$ on both
modes $k_{j}$, the field set $\left\{ \hat{a}_{j},\hat{a}_{j}^{^{\prime
}}\right\} $ is changed into the set $\left\{ \hat{a}_{Rj},\hat{a}%
_{Rj}^{^{\prime }}\right\} $ $(j=1,2)$. \ A general $R(\vartheta ,\;\xi )$
transformation, expressed in terms of complex parameters for which:\ $\left|
\vartheta \right| ^{2}+\left| \zeta \right| ^{2}=1$, relates the two field
sets as follows: 
\begin{equation}
{\widehat{a}_{R1} \atopwithdelims[] \widehat{a}_{R1}^{\prime }}%
=R^{\dagger }%
{\widehat{a}_{1} \atopwithdelims[] \widehat{a}_{1}^{\prime }}%
R=\left[ 
\begin{array}{cc}
\vartheta & \zeta \\ 
-\zeta ^{\ast } & \vartheta ^{\ast }
\end{array}
\right] 
{\widehat{a}_{1} \atopwithdelims[] \widehat{a}_{1}^{\prime }}%
;\ 
{\widehat{a}_{R2}^{\prime } \atopwithdelims[] \widehat{a}_{R2}}%
=R^{\dagger }%
{\widehat{a}_{2}^{\prime } \atopwithdelims[] \widehat{a}_{2}}%
R=\left[ 
\begin{array}{cc}
\vartheta & \zeta \\ 
-\zeta ^{\ast } & \vartheta ^{\ast }
\end{array}
\right] 
{\widehat{a}_{2}^{\prime } \atopwithdelims[] \widehat{a}_{2}}%
\end{equation}

It can be easily checked that Eq. 1 can be re-expressed in terms of the new
field set into the invariant form: $\widehat{H}_{RI}=i\hslash \chi \lbrack 
\widehat{A}_{R}-\widehat{A}_{R}^{\prime }]+h.c.$\ , where $\widehat{A}%
_{R}\equiv $ $\hat{a}_{R1}\hat{a}_{R}$, $\widehat{A}_{R}^{\prime }\equiv $ $%
\hat{a}_{R1}^{\prime }\hat{a}_{R2}^{\prime }$, {\it only} by setting the
OPA\ {\it intrinsic phase} $\Phi =0$, i.e. $\varepsilon =-1$. Interestingly,
note that the same dynamical condition $\Phi =0$ implies the well known $%
SU(2)$ invariance of the $\overrightarrow{\pi }$-entangled$\;$''singlet
states'' generated by Spontaneous Parametric Down Conversion (SPDC), that is
by\ the OPA\ when is not {\it quantum-injected} (or, when it is only
''injected'' by the {\it vacuum field} on both input modes $k_{j}$).

Since, in general the input $N\geq 1$ qubits injected into the amplifier are
quantum superpositions of $\overrightarrow{\pi }$-states, the dynamical
condition $\Phi =0$ finally implies the \ {\it universality} (U) of the
overall cloning and U-NOT\ transformations.\ For the sake of a clearness in
the future discussions, we find convenient to re-cast the invariant
Hamiltonian with $\Phi =0$, in the following form: 
\begin{equation}
\widehat{H}_{int}=i\hslash \chi \left( \widehat{a}_{\pi }\widehat{b}_{\pi
\perp }-\widehat{a}_{\pi \perp }\widehat{b}_{\pi }\right) +h.c.
\end{equation}
where $\widehat{a}$ and\ $\widehat{b}$ are the overall field operators
acting respectively on the output modes $k_{1}$and $k_{2}$. For reasons that
will become clear in the following Sections these modes are referred to as
the {\it cloning}\ (C) and the {\it anticloning}\ (AC) modes, respectively.
\ Furthermore, since $g=\chi t$ is independent of any {\it unknown}
polarization state of the injected field, we have denoted the creation $%
\widehat{a}^{\dagger }$,$\ \widehat{b}^{\dagger }$ and annihilation $%
\widehat{a}$ ,$\ \widehat{b}$ operators of a single photon in modes $k_{1}$
, $k_{2}$ with subscripts $\pi $ or $\pi ^{\perp }$ to indicate the
invariance of the process with respect to the polarization states of the
input particles. Of course, the $SU(2)$ transformation for the fields $%
\widehat{a}$ is again expressed as follows:

\begin{equation}
{\widehat{a}_{R\pi } \atopwithdelims[] \widehat{a}_{R\pi \perp }}%
=R^{\dagger }%
{\widehat{a}_{\pi } \atopwithdelims[] \widehat{a}_{\pi \perp }}%
R=\left[ 
\begin{array}{cc}
\vartheta & \zeta \\ 
-\zeta ^{\ast } & \vartheta ^{\ast }
\end{array}
\right] 
{\widehat{a}_{\pi } \atopwithdelims[] \widehat{a}_{\pi \perp }}%
\end{equation}
and: $\left| \eta \right| ^{2}+$ $\left| \zeta \right| ^{2}=1$. The same
R-transformation \ is valid for the fields $\widehat{b}$. The polarization
conditions $\pi $ and $\pi ^{\perp }$ will be expressed respectively by the
field state vectors $\left| \Psi \right\rangle $ and $\left| \Psi ^{\perp
}\right\rangle $ on the C and AC\ output modes of the apparatus.

\section{Test of the ''universality'' condition}

As already remarked, the {\it universality} (U) condition implies for the
OPA amplifier the $SU(2)$ invariance of $\widehat{H}_{int}$ when the spatial
orientation of the OPA crystal makes it available for SPDC creation of
2-photon entangled {\it singlet} states. Indeed the U-condition in
amplifying physical devices is quite a peculiar property that can only be
realized by a very small number of arrangements set in very special
conditions \cite{20}. Luckily enough the QIOPA\ apparatus has been found to
possess this property. In facts the OPA\ application represents the first
actual realization of the U-condition for $\overrightarrow{\pi }-qubits$ 
\cite{9,13,21}. In spite of the ''microscopic'' quantum theoretical approach
adopted in the previous Section theory we\ should note that in the present
context the universality property is indeed a ''macroscopic'' classical
feature of the OPA device. Thus it can be tested equally well either by
injection of a quasi-classical, e.g. a {\it coherent} (Glauber)\ field, or
by injection of a quantum state, e.g. a Fock state. Because of the relevance
of the U property, we shall undertake the experimental demonstration in both
ways. The ''classical'' test will be described in the present Section while
analogous tests carried out by single-photon Fock states will be reported
later in the paper. In order to do that, let us first venture into a
detailed description of the excitation laser and of the QIOPA apparatus
adopted throughout the present work:\ Figure 3.

{\large Apparatus}. The main source of all experiments reported by this work
was a Ti:Sa mode-locked pulsed laser (Coherent MIRA) providing by Second
Harmonic Generation (SHG) the ''pump'' field for the quantum-injected
optical parametric amplifier QIOPA associated with the spatial mode having
wave-vector (wv) $k_{p}$ and wavelength (wl) $\lambda _{p}=397.5nm$, i.e. in
the ultraviolet (UV) range of the spectrum. The average UV\ power was 0.25
W, the pulse repetition rate was 76 MHz and the time duration of each UV
pulse was $\tau =140$ femtoseconds (fs). The OPA active element, consisting
of a 1.5 mm thick nonlinear (NL) crystal of $\beta $-barium borate (BBO) cut
for Type II phase-matching, was able to generate by SPDC $\overrightarrow{%
\pi }-entangled$ pairs of photons. Precisely, the OPA {\it intrinsic phase}
was set as to generate by SPDC ''singlet'' entangled states on the output
optical modes: $\Phi =0$. The photons of each pair were emitted with equal
wls $\lambda =795nm$ over two spatial modes $k_{1}$ and $k_{2}$ making an
internal angle = 8%
${{}^\circ}$%
. In all experiments the time (t) optical walk-off effects due to the
birefringence of the NL crystal were compensated by inserting in the modes $%
k_{1}$ and $k_{2}$ fixed X-cut, 4.8 mm thick quartz plates. All adopted
photodetectors ($D$), but $D_{2}^{\prime }$\ (Figure 3)\ were equal
SPCM-AQR14 Si-avalanche nonlinear single photon units with nearly equal $%
QE^{\prime }s\cong 0.55$. One interference filter with bandwidth $\Delta
\lambda =6nm$ was placed in front of each detector $D$. Only the detector $%
D_{2}^{\prime }$ was a linear Si photodiode SGD100. Polarizing
beam-splitters (PBS)\ in Figures 3 and 5 were adopted as measurement devices
providing the polarization analysis.

{\large Universality test by ''classical'' field injection}: The U-test was
carried out by injection of the strongly attenuated laser beam, with wl $%
\lambda =795nm,$ contributed via a beam-splitter by the main mode-locked
source and directed along the OPA injection mode, $k_{1}$:\ Figure 3. The
parametric amplification, with ''gain'' $g=0.11$ was detected at the OPA
output mode $k_{2}$ by the linear Si photodiode SGD100 $(D_{2}\prime )$,
filtered by an interference filter with bandwidth $\Delta \lambda =3nm$. The
time t-superposition in the NL crystal of the ''pump'' and of the
''injection'' pulses was assured by micrometric displacements ({\bf Z}) of a
two-mirror optical ''trombone''. The pulse shapes shown by the coincidence
data reported as function of Z in Figure 3, as well as in Figures 4 and 5
later, are indeed the {\it signature for actual amplification}, i.e. arising
from the effective time and space superposition in the NL crystal of the UV
''pump'' pulse and of the optical pulses with $\lambda =795nm$ injected into
the OPA. Different \ $\overrightarrow{\pi }-states$ of the injected field,
formally expressed also by the captions of the inset of Figure 3, were
prepared by a single wave-plate $WP_{1}^{\prime }$, corresponding to a
suitable optical retardation, equal to $\lambda /2$ and $\lambda /4$ between
the two orthogonal basis-states, i.e. horizontal (H) and vertical (V). The
OPA amplified output states were detected by an apparatus inserted on mode $%
k_{2}$ and consisting of the set ($WP_{2}^{\prime }$+ $\overrightarrow{\pi }$%
{\it -analyzer}), the last device being provided by the polarizing beam
splitter $PBS_{2}^{\prime }$.

The universality condition is demonstrated by the plots of Figure 4 showing
the amplification pulses detected by D'2 on the OPA output (AC)\ mode, $%
k_{2} $. Each plot corresponds to a definite state of polarization ($\pi $)
of the field injected on mode $k_{1}$: $[cos(\theta /2)|H>+\exp (i\varphi
)\sin (\theta /2)|V>]$. Precisely, the polarization was set to be either
linear (that is: $\theta $ $=\pi /2,\pi ;\;\varphi =0)$, or circular (that
is: $\theta =\pi /2;\varphi =-\pi /2)$, or very generally elliptical: $%
\theta =5\pi /18,\varphi =-\pi /2$. We may check in Figure 4 that, in spite
of the quite different injected states, the amplification curves are almost
identical. Each coherent pulse injected on the mode $k_{1}$ was amplified
into an average photon number $M^{\prime }=5\times 10^{3}$ on the output
mode $k_{2}$.\newline

\section{Universal Optimal Quantum Cloning Machine and U-NOT\ gate}

The first machine which has been investigated theoretically was the {\it %
optimal universal quantum cloner }of qubits. In the simple case investigated
by the present work, namely of cloning $N=1$ input into $M=2$ output $qubits$%
, the action of the cloner can be described by a simple covariant
transformation \cite{22}:

\begin{equation}
|\Psi >|\downarrow >_{C}|\downarrow >_{AC}\Longrightarrow \sqrt{2/3}|\Psi
>|\Psi >|\Psi ^{\perp }>_{AC}-\sqrt{1/3}|\{\Psi ,\Psi ^{\perp }\}>|\Psi
>_{AC}
\end{equation}
where the first (unknown) state vector in the left-hand side of Equation 13
corresponds to the system to be cloned, the second state vector describes
the system on which the information is going to be copied (''blank'' qubit),
represented by the ''{\it cloning channel}'' (C), the mode $k_{1}$, while
the third state vector represents the cloner. The state $|\Psi ^{\perp }>$
is the {\it antipode} of $|\Psi >$ on the Bloch sphere. In the present work
the ''cloner'' is a qubit, associated with the {\it anti-cloning} channel
(AC). The blank qubit and the cloner are initially in the known state $%
|\downarrow >$. At the output of the cloner we find two qubits (the original
and the copy) in the state $\rho =$ $2/3|\Psi >|\Psi ><\Psi |<\Psi |+$ $%
1/3|\{\Psi ,\Psi ^{\perp }\}><\{\Psi ,\Psi ^{\perp }\}|$ where $|\{\Psi
,\Psi ^{\perp }\}>$ is the completely symmetric state of two qubits. The
density operator $\rho $ describes the best possible approximation of the
perfect cloning process, i.e. the state $|\Psi >|\Psi >$. The fidelity of
this transformation does not depend on the state of the input and reads $F=$ 
$Tr(\rho _{1}\widehat{n}_{1\pi })/Tr(\rho _{1}\widehat{n}_{1})$ $=5/6\approx
0.833$. The cloner itself after the cloning transformation is in the state $%
\rho _{AC}=$ $2/3|\Psi ^{\perp }><\Psi ^{\perp }|+$ $1/3\times {\bf I}${\bf %
\ }, where ${\bf I}$ is the unity operator. This last density operator is
the best possible approximation of the action of the spin-flip (U-NOT)
operation permitted by the quantum mechanics. As we shall see in the next
Sections, for $N=1$ input states the fidelity of the U-NOT transformation is 
$F^{\ast }=Tr(\rho _{2}\widehat{n}_{2\bot })/Tr(\rho _{2}\widehat{n}%
_{2})=2/3\approx 0.666$ and is equal to the fidelity of estimation \cite{4}.
As we shall see, the universal quantum machines have been subsequently
generalized for the case of multi-qubit inputs, when the machines takes as
an input $N>1$ identically prepared qubits and generates either $M>N$ clones
or otherwise transformed qubits.

We may establish a close connection of the above results for ''cloning''
with the adopted {\it universal} QIOPA\ system, by considering for instance
the simple case of a linearly polarized {\it single-photon} $N=1$ injected
into the OPA on the input mode $k_{1}$ with {\it horizontal} (H)
polarization, $\overrightarrow{\pi }=${\bf h}. In virtue of the {\it %
universality} this very particular injection condition indeed expresses the
general behaviour of the amplifier. By the analysis reported in Section 2,
the amplification leads to the output state: $|\Psi >=\widehat{U}|\Psi
>_{IN}=\widehat{U}|1>_{1h}|0>_{1v}|0>_{_{2h}}|0>_{2v}\equiv \widehat{U}%
|1,0,0,0>.$ This last expression implies one photon on the input $%
k_{1h}-mode $ with $\overrightarrow{\pi }=${\bf h} and zero photons on the
three other input modes $k_{1v}$, $k_{2h}$, $k_{2v}$. Let's adopt the
isomorphism: $|\Psi >|\downarrow >=|1>_{1h}|0>_{1v}$, $|\Psi ^{\perp
}>|\downarrow >=|0>_{1h}|1>_{1v}$and $|\{\Psi ,\Psi ^{\perp
}\}>=|1>_{1h}|1>_{1v}$, we shall find that the output state of the OPA\
amplifier $|\Psi >_{out}$expressed by Equation 5 is identical, in the first
order of the small parameter $\Gamma $, to the output of the cloner
transformation expressed by Equation 13. Indeed the input mode $\left|
1,0,0,0\right\rangle \ $evolves into the first order state: 
\begin{equation}
\widehat{U}\left| 1,0,0,0\right\rangle \approx \left| 1,0,0,0\right\rangle
-\Gamma \left( \sqrt{2}\left| 2,0\right\rangle _{k_{1}}\otimes \left|
0,1\right\rangle _{k_{2}}-\left| 1,1\right\rangle _{k_{1}}\otimes \left|
1,0\right\rangle _{k_{2}}\right)
\end{equation}
where for clarity we set: $\left| x,\ y\right\rangle _{k_{j}}\equiv \left|
x\right\rangle _{jh}\otimes \left| y\right\rangle _{jv}$ and $j=1,2$. The
approximation for the state vector describing the two $k_{j}$-$%
\mathop{\rm mod}%
$es at times $t>0$ is sufficient in all cases in which the coupling value is
small, as in our experiment: $g\thicksim 0.11$. The two addenda at the
right-hand side of Equation 14 represent respectively the process when the
input photon in the mode $k_{1}$ did not interact in the nonlinear medium,
followed by the first order amplification process in the OPA. Here the state 
$\left| 2,0\right\rangle _{k_{1}}$ describing two photons of the {\it cloning%
}\ mode $k_{1}$ in the polarization state $\overrightarrow{\pi }$
corresponds to the state $\left| \Psi \Psi \right\rangle $. This
state-vector describes the $1\rightarrow 2\ {\it cloning}$ of the original
injected photon. The vector $\left| 0,1\right\rangle _{k_{2}}$ describes the
state of the {\it anticloning}\ mode $k_{2}$ with a single photon with the
polarization $\pi ^{\perp }$ corresponding to the state $|\Psi ^{\perp
}>_{AC}$. This state vector represents the {\it flipped} version of the
input. To see that the stimulated emission is indeed responsible for
creation of the flipped qubit, let us compare the result above with the
output of the OPA when the vacuum is injected into the nonlinear crystal,
i.e. the SPDC\ process. In this case, we obtain to the same order of
approximation: 
\begin{equation}
\widehat{U}\left| 0,0,0,0\right\rangle \approx \left| 0,0,0,0\right\rangle
_{k_{1}}-\Gamma \left( \left| 1,0\right\rangle _{k_{1}}\otimes \left|
0,1\right\rangle _{k_{2}}-\left| 0,1\right\rangle _{k_{1}}\otimes \left|
1,0\right\rangle _{k_{2}}\right)
\end{equation}
\newline
We see that the flipped qubit described by the state vector $\left|
0,1\right\rangle _{k_{2}}$ in the right-hand sides of Equations 14 and 15 do
appear with different amplitudes corresponding to the ratio of the
probabilities to be equal to: $R=2:1$. Our experiment, which deals with the
injection of $N=1$ photon, indeed consists of the measurement of $R$, as we
shall see. According to the analysis reported in Section 2 the cloning and
the U-NOT operations are not altered by the multiplication of $\widehat{H}%
_{I}$ by any overall phase factor.

As anticipated in Section 1, it is well known that a method for flipping
qubits alternative to the present U-NOT gate consists on manufacturing the
orthogonal qubits on the basis of the result of the measurement on the input
qubits. The fidelity $F^{\ast }\ $of the two alternative methods is indeed
the same \cite{3,4}. However, in the U-NOT gate the information encoded in
the input qubit is not lost in the irreversible state reduction implied by
the measurement. It is just redistributed into several qubits at the output.
Since this information redistribution is governed by a unitary
transformation, the process is in principle reversible which is definitely
not true in the case of measurement-based flipping operation. We also want
to stress that, to the best of our knowledge the recent U-NOT experiment
reported by us in \cite{9}, and the present one are the first systematic
attempts to realize an {\it anti-unitary} gate. Obviously, the fidelity of
the gate is strictly determined by the structure, i.e. rules of quantum
mechanics, as shown. This stresses the fundamental importance to understand
the action of the U-NOT gate in view of a clarification of the structure of
quantum mechanics in connection with the CP-map topology, as we shall see 
\cite{1}.

A more general analysis can be undertaken by extending the isomorphism
discussed above to a larger number of input and output particles $N$ $\ $and 
$M$. In this case too it is found that the QIOPA\ amplification process $%
\widehat{U}_{AA^{\prime }}$ in each order of the decomposition into the
parameter $\Gamma $ corresponds to the $N\longrightarrow M\;$cloning
process. Precisely, in this case $M\geq N$ output particles are detected
over the output {\it cloning}\ (C) mode, $k_{1}$. Correspondingly, $M-N$
particles are detected over the output {\it anticloning} (AC) mode, $k_{2}$.
Indeed the {\it optimum cloning} output state determined by the theory of
the {\it quantum-injected} OPA is found 
\begin{equation}
|\Psi _{N}>_{M}=\sum 
{\textstyle{M-N \atop m=0}}%
(-1)^{m}P%
{\textstyle{M \atop N}}%
(m)|M-m,\ m,\ m,\ M-N-m>
\end{equation}
\newline
where: $P%
{\textstyle{M \atop N}}%
$ $(m)=\left[ 
{M-m \choose N}%
/%
{M+1 \choose N+1}%
\right] ^{1/2}$. By a simple re-arrangement of the Fock state degeneracies,
Equation 16 is found to agree exactly with the general expression of the
output state of the QIOPA given by Equation 5 \cite{13,19}. In this general
case {\it optimal cloning fidelity }is found \cite{23,24}: $%
F=(NM+M+N)/(MN+2M)$. For $N=1$, $M=2$ is $F=5/6\approx 0.833$. It appears
clear from the above analysis that the effect of the input {\it vacuum field,%
} which is {\it necessarily} injected in any {\it universal} optical
amplifier, is indeed responsible to reduce the ''fidelity'' of the quantum
machines at hand. More generally the vacuum field is in exact correspondence
with, and must be interpreted as the amount of quantum fluctuations that
determines the upper bounds to the {\it Fidelity} determined by the very
CP-map structure of quantum mechanics.

At last note that, luckily enough, the QIOPA apparatus is found to be an
ideal system to demonstrate, in a peculiar unifying manner, the relevant
features of the most interesting machines investigated thus far:\ the UOQCM
and the U-NOT gate. The present work also establishes an interesting
connection between the technical {\it engineering} of parametric amplifiers
and the abstract quantum measurement theory.

\section{Experimental UOQCM and U-NOT gate}

The main laser apparatus and the basic structure of the NL\ parametric
amplifier was already described in Section 3. Figure 5 shows the QIOPA
apparatus, arranged in a {\it self-injected} configuration and adopted to
realize {\it simultaneously} the Universal Optimal Quantum Cloning Machine
(UOQCM) and the Universal Optimal NOT gate. The UV pump beam, back reflected
by a spherical mirror $M_{p}$ with 100\% reflectivity and micrometrically
adjustable position ${\bf Z}$, excited the NL OPA crystal amplifier in both
directions $-k_{p}$ and $k_{p}$, i.e. correspondingly oriented towards the
Right (R) and the Left (L) sides of \ the figure. A SPDC process excited by
the $-k_{p}$ pump mode created single pairs of photons with equal wls $%
\lambda =795nm$ in entangled singlet-states of linear polarization, $%
\overrightarrow{\pi }$. One photon of each pair, emitted over $-k_{1}$ was
reflected by a spherical mirror $M$ into the NL crystal where it provided
the $N=1$ {\it quantum injection} into the OPA excited by the UV-pump beam
associated with the back reflected mode $k_{p}$. Because of the low pump
intensity, the probability of the unwanted $N=2$ photon injection has been
evaluated to be $10^{-2}$ smaller than the one for $N=1$. The twin SPDC\
generated photon emitted over $-k_{2}$ was selected by the devices
(Wave-Plate + Polarizing Beam Splitter: $WP_{T}\ $+ $PBS_{T}$), detected by $%
D_{T}$ and provided the ''trigger'' of the overall conditional experiment.
Because of the EPR non-locality implied by the SPDC\ emitted singlet state,
the selection on mode $-k_{2}$ provided the realization on $k_{1}$ of the
polarization states $\left| \Psi \right\rangle _{IN}$ of the injected
photon. By adopting $\lambda /2$ or $\lambda /4$ Wave-Plates (WP) with
different orientations of the optical axis, the following states were
injected:$\;\left| H\right\rangle $, $2^{-1/2}(\left| H\right\rangle +\left|
V\right\rangle )$, and $2^{-1/2}(\left| H\right\rangle +i\left|
V\right\rangle )$. The three fixed quartz plates $(Q)$ inserted on the modes 
$k_{1}$, $k_{2}$ and $-k_{2}$ provided the compensation for the unwanted
walk-off effects due to the birefringence of the NL crystal. An additional
walk-off compensation into the BBO\ crystal was provided by the $\lambda /4$
WP exchanging on the mode $-k_{1}\ $the $\left| H\right\rangle $ and $\left|
V\right\rangle $ $\overrightarrow{\pi }$ components of the injected photon.

As we shall see, the goal of the present {\it cloning} experiment was to
measure, under injection of the state $\left| \Psi \right\rangle $, the OPA\
amplification $R\;$on the output C\ mode carrying the {\it same} $%
\overrightarrow{\pi }$ condition corresponding of the input state.
Contextually, no amplification should affect the output state corresponding
to the polarization $\overrightarrow{\pi }^{\perp }$ orthogonal\ to $%
\overrightarrow{\pi }$. In order to perform this task, the $PBS_{2}$ was
removed on the mode $k_{2}$ and the photons on the same mode detected by a
single detector: $D_{2}$. The two-cloning photons associated with\ the C
mode, $k_{1}$ were separated by means of a $50:50$ conventional Beam
Splitter (BS)\ and their polarization states were analyzed by the
combinations of $WP_{1}$ and of $PBS_{1a}$ and $PBS_{1b}$. For each injected
state of polarizations $\left| \Psi \right\rangle ,$ $WP_{1}$ was set in
order to detect $\left| \Psi \right\rangle $ by detectors $D_{a}$ and $D_{b}$
and to detect the state orthogonal to $\left| \Psi \right\rangle $, viz. $%
\left| \Psi ^{\perp }\right\rangle $ by $D_{b}^{\ast }.$ Hence any
coincidence event detected by $D_{a}$ and $D_{b}$ corresponded to the
realization of the state $\left| \Psi \Psi \right\rangle _{1}$ over the C
mode, while a coincidence detected by $D_{a}$ and $D_{b}^{\ast }$
corresponded to the state $\left| \Psi \Psi ^{\perp }\right\rangle _{1}$.

The measurement of $R$ was carried out by 4-coincidence measurements
involving simultaneously the detectors of the set $[D_{2},D_{T},D_{a},D_{b}]$
and the one of the set: $[D_{2},D_{T},D_{a},D_{b}^{\ast }]$. The adopted
4-coincidence electronic apparatus had a time resolution $\tau =3$ nsec.

The experimental data reported\ on the left column of Figure 6 correspond to
the 4-coincidence measurement by $[D_{2},D_{T},D_{a},D_{b}],$ that is, to
the emission over the C-mode of the state $\left| \Psi \Psi \right\rangle
_{1}$ under injection of the input state $\left| \Psi \right\rangle _{IN}$.
As for all experiments reported in Figures 4, 6, 7, the resonance peaks
found by this last measurement identified the position $Z$ of the UV\
retro-reflecting mirror for which the spatial and temporal superposition of
the UV-pump pulses and of the injected single photon pulses was realized. In
other words the peaks identified the actual realization of the {\it quantum
injected} amplification of the OPA\ apparatus. The right hand side (r.h.s.)
column of Figure 5 reports the corresponding 4-coincidence data obtained by
the set $[D_{2},D_{T},D_{a},D_{b}^{\ast }]$, i.e. implying the realization
on the C mode of the state $\left| \Psi \Psi ^{\perp }\right\rangle _{1}$.
As expected from theory no amplification peaks are present in this case,
here referred to as ''noise''. The value of the signal-to-noise ratio $R$
determined by the data shown in the left and right sides of Figure 6 in
correspondence with each injection condition $\left| \Psi \right\rangle
_{IN} $ was adopted to determine the corresponding value of the {\it cloning
fidelity\ }$F=Tr(\rho _{1}\widehat{n}_{1\pi })/Tr(\rho _{1}\widehat{n}_{1})$=%
$(2R+1)/(2R+2)$\ implied by Eqs. 14, 15 with the definitions: $\widehat{n}%
_{1}=\widehat{n}_{1\pi }+\widehat{n}_{1\bot }$, $\widehat{n}_{1\pi }\equiv 
\hat{a}_{\pi }^{\dagger }\hat{a}_{\pi }$, $\widehat{n}_{1\bot }\equiv \hat{a}%
_{\pi \bot }^{\dagger }\hat{a}_{\pi \bot }$. Precisely, the values of the
signal-to-noise ratio $R\;$ were determined as the ratio of the peak values
of the plots on the left hand side (l.h.s.)\ of Figure 6 and of the values
of the corresponding plots on the r.h.s. measured at the same value of ${\bf %
Z}$. The results are: $F_{H}=0.812\pm 0.007$;$\ F_{H+V}=0.812\pm 0.006$; $%
F_{left}=0.800\pm 0.007$, to be compared with the theoretical ''optimal''
value: $F_{th}=5/6\approx 0.833$ corresponding to the limit value of the
amplification ratio: $R=2:1$ \cite{21,25}.

The U-NOT\ gate operation has been demonstrated by restoring the $PBS_{2}$
on the AC-mode, $k_{2}$ coupled to the detectors $D_{2}$, $D_{2^{\ast }}$,
via the $WP_{2}$, as shown in Figure 5. The $\overrightarrow{\pi }\;$%
-analyzer consisting of $(PBS_{2}+WP_{2})$ is set as to reproduce the {\it %
same} filtering action of the analyzer $(PBS_{T}+WP_{T})$ for the
''trigger'' signal. On the C channel, $k_{1}$, the devices $PBS_{1a}$ and $%
PBS_{1b}$ were removed and the field was simply coupled by the {\it normal }%
Beam Splitter $(BS_{1})\;$to the detectors $D_{a}$ and $D_{b}$. A
coincidence event involving these ones was the {\it signature} for a {\it %
cloning} event. The values of the signal-to-noise ratio $R^{\ast }$
evaluated on the basis of the data of the 4-coincidence experiments
involving the sets $[D_{2},D_{T},D_{a},D_{b}]$ and $[D_{2^{\ast
}},D_{T},D_{a},D_{b}]$ and reported in Figure 7, were adopted to determine
the corresponding values of the U-NOT {\it fidelity} $F^{\ast }=Tr(\rho _{2}%
\widehat{n}_{2\bot })/Tr(\rho _{2}\widehat{n}_{2})=R^{\ast }/(R^{\ast }+1)$
implied by Eqs 14, 15 and: $\widehat{n}_{2}=\widehat{n}_{2\pi }+\widehat{n}%
_{2\bot }$, $\widehat{n}_{2\pi }\equiv \widehat{b}_{\pi }^{\dagger }\widehat{%
b}_{\pi }$, $\widehat{n}_{2\bot }\equiv \widehat{b}_{\pi \bot }^{\dagger }%
\widehat{b}_{\pi \bot }$. The results are: $\ F_{H}^{\ast }=0.630\pm 0.008$;$%
\ F_{H+V}^{\ast }=0.625\pm 0.016$;$F_{left}^{\ast }=0.618\pm 0.013$ to be
compared with the theoretical ''optimal'' value: $F^{\ast }=2/3\approx 0.666$
again corresponding to the limit value: $R=2:1$. The 4-coincidence method
was adopted in all experiments described in this Section because it allowed
a better\ spatial mode filtering of the system, leading to a larger
experimental value of $R$. A 3-coincidence cloning experiment involving the
sets $[D_{2},D_{T},D_{b}]$ and $[D_{2},D_{T},D_{b}^{\ast }]$ was also
carried out successfully but\ attaining a small amplification ratio: $%
R=1.18:1$. An analogous result was obtained by a 3-coincidence U-NOT\ gate
experiment. Note also that the height of the amplification peak in Figures
6,7 does not decrease towards zero for large values of ${\bf Z}$, as
expected. This effect is attributable entirely to the limited time
resolution of the 4-coincidence apparatus. The effect would disappear if the
resolution could be pushed into the sub-picosecond range, precisely of the
order of the time duration of the interacting pump and injection pulses: $%
\tau ^{\prime }\simeq 140\ $f$\sec $. Such a resolution is hardly obtainable
with the present technology.

{\large Universality test by quantum injection. } As concluding remark, note
that all experimental results reported in the left columns of Figures 6 and
7 show an amplification efficiency, viz a value of the signal-to-noise ratio 
$R$ which is almost identical for the adopted different $\overrightarrow{\pi 
}$ conditions corresponding to the input state:\ $\left| \Psi \right\rangle
_{IN}$. This very significant result represents the first demonstration of
the ''universality'' of the QIOPA\ system carried out by {\it quantum
injection} of a {\it single} qubit. It is alternative to the \ {\it %
classical field} \ universality test reported earlier in Section 3 and in 
\cite{9,21}.

\section{Interpretation\ \ \ \ \ }

A remarkable, somewhat intriguing aspect of the present work is that both
processes of quantum cloning and the U-NOT gate are realized contextually by
the {\it same} physical apparatus, by the same unitary transformation and
correspondingly by the same quantum logic network. To the best of our
knowledge it is not well understood yet why these ''{\it forbidden'' }%
processes can be so closely related.\thinspace We may try to enlighten here
at least one formal aspect of this correlation.

Remind first that the two processes are detected in our experiment over the
two $\overrightarrow{\pi }-${\it entangled} output (C) and (AC) channels,
physically represented by the two corresponding wavevectors $\overrightarrow{%
k_{j}}$ $(j=1,2)$. In addition, the overall output vector state $\left| \Psi
\right\rangle $ given by Eq.\ 5 is a {\it pure state} since the {\it unitary}
evolution operator $\widehat{U}=\widehat{U}_{A}\otimes \widehat{U}%
_{A^{\prime }}$ (Eq. 3) acts on a {\it pure} input state. As a consequence,
the reduced density matrices $\rho _{1}\ $and $\rho _{2}$ have the same
eigenvalues \cite{26} and the entanglement of the {\it bi-partite }state $%
\left| \Psi \right\rangle \ $can be conveniently measured by the {\it %
entropy of entanglement:} 
\begin{equation}
E(\Psi )=S(\rho _{1})=S(\rho _{2})
\end{equation}
where $S(\rho _{j})=-Tr\rho _{j}\log _{2}\rho _{j}$ is the Von Neumann
entropy of the either (C) or (AC) subsystem: $j\ (j=1,2)\ $\cite{27}. Let's
consider the approximate {\it cloning} process performed by our {\it %
universal optimal machine} (UOQCM)\ acting on the C channel, $k_{1}$. What
has been actually realized in the experiment was a procedure of
''linearization'' of the cloning map which is {\it nonlinear }and, as such%
{\it \ }cannot be{\it \ }realized exactly by Nature \cite{1,5,6}. By this
procedure a {\it mixed-state} condition of the output state$\ \rho _{1}$ was
achieved corresponding to the {\it optimal} limit value of the entropy $%
S(\rho _{1})>0$. In virtue of Eq.\ 17 the {\it same} {\it mixing} condition,
with the {\it same} entropy $S(\rho _{2})\ $is shared by the state $\rho
_{2} $ on the AC channel leading to the {\it noisy} measurement process for
the {\it flipped-qubit} $|\Psi ^{\perp }>$. Note also that the present
argument relies {\it only} on the {\it map-linearity}\ without accounting
explicitly for the {\it non} {\it complete-positivity} of the time reversal
map $T$, the one which is actually responsible for the flipping dynamics 
\cite{3}. The presence of the quoted {\it unavoidable} quantum noise is
shown by the experimental plots in the r.h.s. of Fig. 7. Furthermore, in
Section 5 is shown that the measurement process implied by the investigation
on qubit-flipping on the AC-channel parallels almost identically the one
realized to investigate cloning on the C-channel.

Very interestingly the above argument could be {\it reversed}, as the
mixed-state condition $S(\rho _{2})>0$ imposed by the $T$ map on the AC
channel is transferred to the C channel, in virtue of Eq. 17, and accounts
for the unavoidable quantum noise affecting the cloning results shown in
Fig. 6. In summary, Equation 17 appears to establish, in the context of the
present experiment an unexpected, perhaps deeply significant {\it symmetry
condition} between two distinct, fundamental constituents of axiomatic
quantum theory \cite{1,28}.

\ Note in this connection that the adoption of the{\it \ classical}\ OPA\
apparatus within an investigation pushed at the extreme quantum measurement
limits, enlightened the conceptual significance of the quantum noise,
originating in this case from the QED\ vacuum field. We remind here that the
failure of an old proposal for\ superluminal signaling by an {\it amplified}
EPR scheme was attributed to the impossibility of realizing a single photon 
{\it ideal} amplifier, i.e.one for which the signal-to-noise ratio would be: 
$R>2$ \cite{6,20}. \newline
For the sake of further clarification, let's analyze the simple case of a 
{\it mode-degenerate} QIOPA\ amplifier in which {\it only} two e.m. modes, $%
k_{\shortparallel }$and $k_{\perp }$ with the same wavelength $\lambda
=2\lambda _{p}$, interact in the NL\ crystal. In work \cite{29} this case
has been realized by a Type II BBO\ crystal cut for {\it collinear} emission
over a {\it single} wavevector $\overrightarrow{k}$ and the modes $%
k_{\shortparallel }$and $k_{\intercal }\ $express two orthogonal states of
polarization, e.g. of linear $(\overrightarrow{\pi })$. This condition is
still represented by the device shown in Figure 2 in which the modes $k_{1}$%
and $k_{2}$ are made to collapse into a single $k$.\ In general the {\it %
quantum injection} is still provided there by a qubit encoded in a single
photon: $\left| \Psi \right\rangle _{IN}\equiv (\widetilde{\alpha }\left|
1,0\right\rangle +\widetilde{\beta }\left| 0,1\right\rangle )$\ where again$%
\ \left| \widetilde{\alpha }\right| ^{2}+\left| \widetilde{\beta }\right|
^{2}=1$ and: $\left| x,y\right\rangle \equiv \left| x\right\rangle
_{\shortparallel }\otimes \left| y\right\rangle _{\perp }$. For the present
purpose the problem may be simplified, with no loss of generality by
assuming: $\widetilde{\alpha }=1$, $\widetilde{\beta }=0$. Applying the
theory given in Section 2, we get the output state to the first order of
approximation: 
\begin{equation}
\left| \Psi \right\rangle \approx \left| 1,0\right\rangle -\sqrt{2}\Gamma
\left| 2,1\right\rangle
\end{equation}
to be compared with Eq.14. Likewise, the no-injection condition leads to the
output state $\;\left| \Psi \right\rangle _{0}\approx \left|
0,0\right\rangle -\Gamma \left| 1,1\right\rangle $, to be compared with
Eq.15. The ''no-interaction'' first term is easily discarded by a
coincidence measurement \cite{29}. \ The second term consists of a{\it \
pure }state realizing both \ {\it exact} cloning and {\it qubit-flipping} on
the output modes. Thus means that any test aimed at the detection of the two
processes {\it is not affected} by quantum noise. However, the device works 
{\it just }for {\it one particular} input qubit, i.e.\ for one particular
choice of the parameters $\widetilde{\alpha }$ and $\widetilde{\beta }$. Any
other choice would lead to a different amplification quantum efficiency. In
other words, the device is not a {\it universal machine}. In analogy with
the theoretical results of Section 2, this may be attributed to the fact the
interaction hamiltonian of the present OPA\ system $\widehat{H}_{I}=i\hslash
\chi \hat{A}+h.c.$, $\hat{A}\equiv $ $\hat{a}_{\shortparallel }\hat{a}%
_{\perp }$ cannot be made formally invariant under displacements of the
injected qubit over the entire Bloch sphere, Figure 1. Indeed at least four
nonlinearly interacting e.m. modes are needed to attain such invariance of
the optical amplifier/squeezing hamiltonian. All these results are fully
consistent with the general theory of the ''optimal quantum machines''\cite
{4,22}.

In summary, the present work shows experimentally in a unifying manner, and
perhaps for the first time, that the concepts of \ no-signaling, i.e.
causality, linearity and complete-positivity have yet little explored
inter-connections within the inner structure of quantum mechanics \cite{6,9}%
. We do believe that the actual results, the suggestions and the open
problems contributed by the present work could be useful by setting
measurement bounds and fundamental performance limitations in the domain of
Quantum Information. Furthermore they should somewhat contribute to a
clarification of some deep structural aspects of axiomatic quantum theory.

We are indebted with Vladimir Buzek, Sandu Popescu, Christoph Simon and
Mauro D'Ariano for enlightening discussions and suggestions. This work has
been supported by the FET European Network on Quantum Information and
Communication (Contract IST-2000-29681: ATESIT), by PRA-INFM\ ''CLON''\ and
by PAIS-INFM\ 2002 (QEUPCO).

\centerline{\bf Figure Captions}

\vskip 8mm

\parindent=0pt

\parskip=3mm

Figure.1. \ Bloch sphere, the state space of a quantum-bit ({\it qubit}). 
{\it Pure-states} are represented by points on the sphere while {\it %
statistical-mixtures} are points inside the sphere. The universal-NOT gate
operation corresponds to the inversion of the sphere, i.e. states $\left|
\Psi \right\rangle $ and $\left| \Psi ^{\bot }\right\rangle $ are antipodes.

Figure.2. \ Schematic diagram of the {\it Quantum Injected} Optical
Parametric Amplifier (QIOPA) in {\it entangled configuration}. The injection
is provided by an external Spontaneous Parametric Down Conversion source
(SPDC) of polarization $(\overrightarrow{\pi })$ entangled photon states.

Figure.3. \ Injection apparatus of attenuated coherent optical pulses into a
QIOPA to demonstrate the universality of the parametric amplification.

Figure.4. Experimental verification of the universality of the OPA system by
injection of $\ \overrightarrow{\pi }-qubits$ represented by several
significant points on the Bloch sphere. The injected state is encoded in a
very attenuated coherent laser beam.

Figure.5. Schematic diagram of the {\it universal optimal cloning machine}
(UOQCM)\ realized on the cloning (C)\ channel (mode $k_{1}$) of a {\it %
self-injected} OPA and of the Universal NOT (U-NOT) gate realized on the
anticloning (AC)\ channel, $k_{2}$. Micrometric adjustments of the
coordinate {\bf Z} of the UV\ mirror $M_{p}$ ensured the time superposition
in the active NL\ crystal of the UV\ 150 femtosecond pump pulses and of the
single photon pulse injected via back reflection by the fixed\ mirror $M$.

Figure.6. \ Demonstration of the UOQCM by single photon qubit self-injection
and 4-detector coincidences. The plots on the left-hand-side (l.h.s) and on
the r.h.s of the figure mutually correspond. The values of the ''{\it %
fidelity}'' of the process $F$ evaluated by each test are expressed in the
upper side of the l.h.s.\ plots while the corresponding test qubits are
expressed in r.h.s. plots.

Figure.7. Demonstration of the Universal Optimal NOT-gate by single photon
qubit self-injection and 4-detector coincidences. The plots on the l.h.s.
and on the r.h.s of the figure mutually correspond. The values of the
fidelity of the process, $F^{\ast }$and the expression of the corresponding
test qubits are shown as for figure 6.

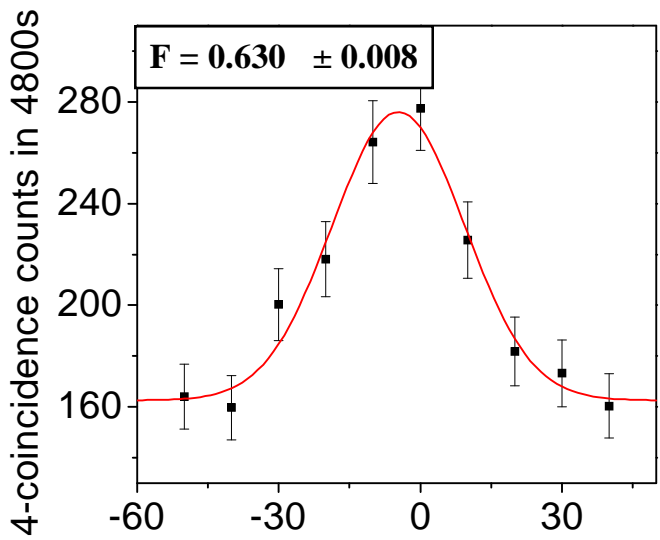
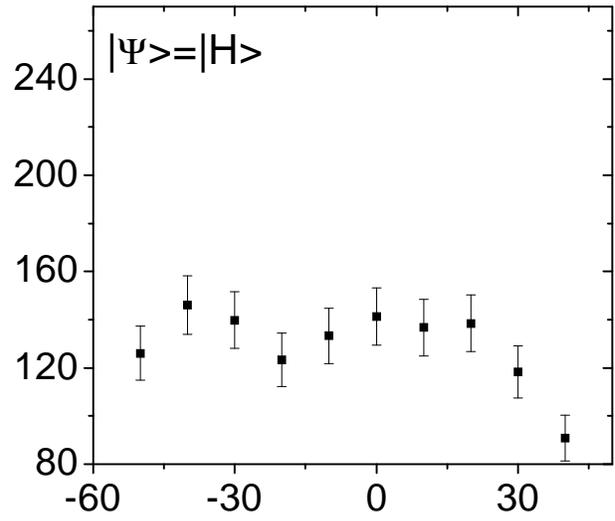
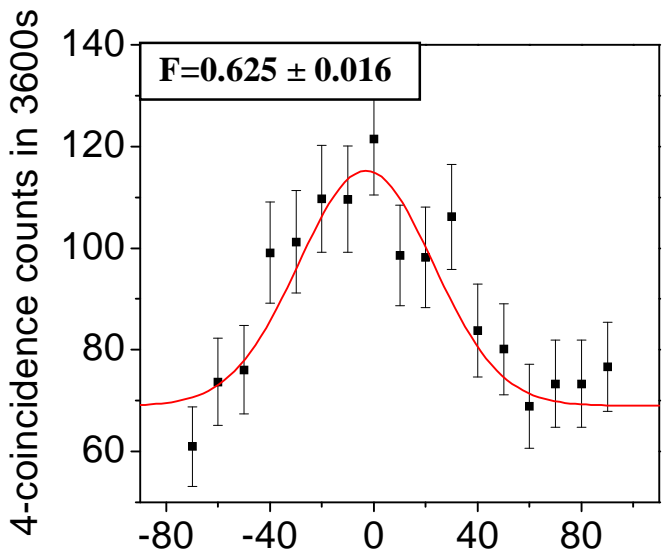
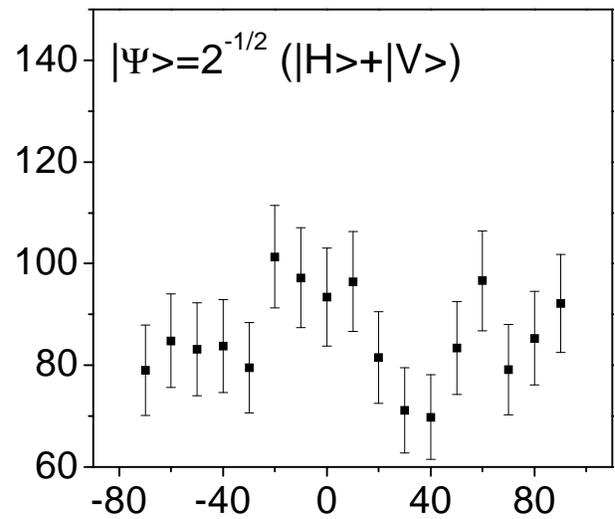
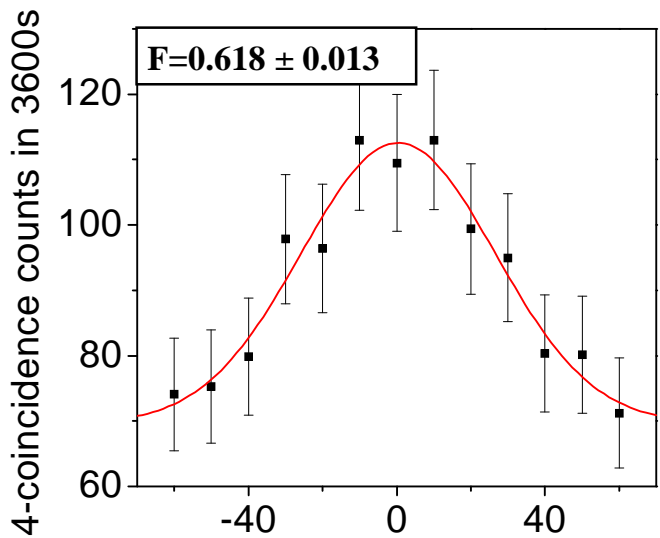
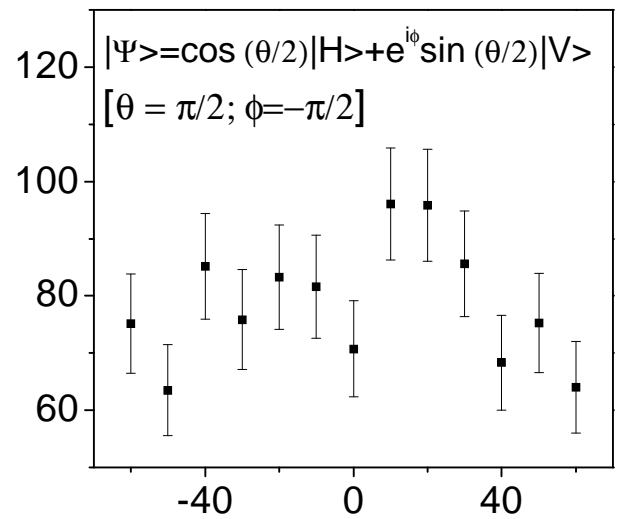
\end{document}